\documentclass[12pt]{article}
\usepackage{epsfig,amsfonts,amssymb}
\usepackage{hyperref}
\usepackage{cite}
\input epsf.sty
\usepackage{cite}

\def\ZZZ{{\hbox{ Z\kern-1.6mm Z}}}
\def\RRR{{\hbox{ R\kern-2.4mm R}}}
\def\CCC{{\hbox{ C\kern-2.3mm C}}}
\def\zzz{{\hbox{z\kern-1mm z}}}

\newcommand{\qeq}{{\hbox{=\kern-2.3mm ? \kern.5mm }}}
\renewcommand{\qeq}{=}

\newcommand{\eps}{\epsilon}

\newcommand{\AAA}{{\cal A}}

\newcommand{\WWW}{{\cal W}}
\newcommand{\VVV}{{\cal V}}

\newcommand{\wt}{\widetilde}
\newcommand{\wh}{\widehat}

\newcommand{\NN}{{\cal N}}

\newcommand{\be}{\begin{equation}}
\newcommand{\ee}{\end{equation}}
\newcommand{\ben}{\begin{eqnarray}\displaystyle}
\newcommand{\een}{\end{eqnarray}}

\newcommand{\refb}[1]{(\ref{#1})}
\newcommand{\p}{\partial}
\newcommand{\sectiono}[1]{\section{#1}\setcounter{equation}{0}}

\def\one{{\hbox{ 1\kern-.8mm l}}}
\def\zero{{\hbox{ 0\kern-1.5mm 0}}}

\newcommand{\bea}[1]{\begin{eqnarray}\label{#1} }
\newcommand{\eea}{\end{eqnarray}}

\begin{document}

~\hfill HRI/ST/1208, ICTS/2012/09

\baselineskip 24pt

\begin{center}
{\Large \bf Black Hole Bound State Metamorphosis }

\end{center}

\vskip .6cm
\medskip

\vspace*{4.0ex}

\baselineskip=18pt

\centerline{\large \rm Abhishek Chowdhury$^a$, Shailesh Lal$^b$, 
Arunabha Saha$^a$
and Ashoke Sen$^a$}

\vspace*{4.0ex}

\centerline{\large \it $^a$Harish-Chandra Research Institute}
\centerline{\large \it  Chhatnag Road, Jhusi,
Allahabad 211019, India}
\centerline{\large \it $^b$International Centre for Theoretical Sciences -- TIFR}
\centerline{\large \it  TIFR Centre Building, Indian Institute of Science}
\centerline{\large \it  Bangalore, India 560012}

\vspace*{1.0ex}
\centerline{\small E-mail:  abhishek@mri.ernet.in, shailesh.lal@icts.res.in,
arunabha@hri.res.in, 
sen@mri.ernet.in}

\vspace*{5.0ex}

\centerline{\bf Abstract} \bigskip

$\NN=4$ supersymmetric string theories contain negative discriminant
states whose numbers are known precisely from microscopic counting
formul\ae. On the macroscopic side,
these results can be reproduced by 
regarding these states as multi-centered black hole
configurations provided we make
certain identification of apparently distinct multi-centered
black hole configurations according to a precise set of rules. In this paper we
provide a physical explanation of  such identifications, thereby establishing that
multi-centered black hole configurations reproduce correctly the microscopic
results for the number of negative discriminant states without any ad hoc
assumption. 

\vfill \eject

\baselineskip=18pt

\tableofcontents

\sectiono{Introduction} \label{sint}

Matching of microscopic counting of BPS states to the entropy of 
supersymmetric black holes is an important problem. Exact
microscopic counting of BPS states,  including the dependence of the spectrum 
on the
asymptotic moduli, has now been
achieved for a wide class of states in $\NN=8$ supersymmetric string
theories\cite{9903163,0506151,0803.1014} and a 
wide class of $\NN=4$ supersymmetric 
string theories\cite{9607026,0412287,0505094,0506249,0508174,0510147,0605210,
0607155,0609109,0802.0544,0802.1556,0803.2692,0702141,0702150,0706.2363} 
in four dimensions. 
An important class
of these microscopic 
states are the so called negative discriminant states -- states carrying
charges for which there are no classical supersymmetric 
single centered black holes but
whose microscopic index is nevertheless non-zero. In particular such states
are abundant in $\NN=4$ supersymmetric string theories. It was shown in
\cite{1104.1498}, following an earlier observation of \cite{0702150}, that all the known 
negative discriminant
states in $\NN=4$ supersymmetric string theories, 
which appear in the microscopic counting of states, can be accounted for precisely
as 2-centered black hole 
configurations, with each center representing a small half-BPS black
hole. This however required one crucial assumption: certain 2-centered
configurations, whose indices can be computed and shown to be the same, had
to be treated as identical configurations. This identification was ad hoc,
since the configurations which had to be identified 
appeared to be different configurations carrying the same total charge. Nevertheless
\cite{1104.1498} gave a
precise set of rules for determining when a pair of configurations have
to be identified. This phenomenon was called black hole metamorphosis. A similar
phenomenon in the context of supersymmetric gauge theories had been discussed
earlier in \cite{0712.3625}. 

Our main goal in this paper will be to understand the physical
origin of this phenomenon, and justify the ad hoc prescription of \cite{1104.1498} for
identifying certain apparently different configurations of black holes.
What we shall show is that precisely for the class of two centered solutions
for which the ad hoc identification rule is to be imposed, one of the black hole
centers need to be replaced by a smooth gauge theory dyon to avoid
certain singularities in the solution. The effect of this is that 
the range of
moduli for which each solution exists is smaller than the one based on the naive
analysis of a two centered black hole solution. Taking into account this
effect, we find that at any given point of the moduli space the total index
contributed by all the two centered configurations which exist at that point adds up
to match precisely the microscopic result for the same index.
Although we have carried out our analysis in the context of a specific 
theory -- for heterotic string theory on $T^6$ -- and worked in a region of the moduli
space where one of the two centers is light and can be regarded as a test particle
in the background produced by the other center, we expect that our analysis
captures the essential physics behind the phenomenon of black hole bound state
metamorphosis for  more general theories and in generic region of the moduli
space.

\sectiono{Review of black hole metamorphosis} \label{smeta}

In this section we shall review the phenomenon of black hole bound state
metamorphosis discussed in \cite{1104.1498}.
Although this phenomenon takes place in all
$\NN=4$ supersymmetric string theories, we shall consider in this paper the
concrete example of heterotic string theory compactified on $T^6$. 
At a generic
point in the moduli space this theory
has 28 U(1) gauge fields and hence a BPS state is characterized by a 28
dimensional electric charge vector $Q$ and a 28 dimensional magnetic charge
vector $P$. We shall denote the combined charge vector as
$(Q, P)$. We can associate with these vectors 
T-duality invariant bilinears $Q^2$,
$P^2$ and $Q\cdot P$. We consider quarter BPS states carrying charges
$(\wh Q, \wh P)$   satisfying
\be \label{egcd}
(\wh Q^2 \wh P^2 - (\wh Q\cdot \wh P)^2) <0, \quad \hbox{and} \quad
\gcd \{\wh Q_i \wh P_j - \wh Q_j \wh P_i, \quad 1\le i,j\le 28\} = 1\, .
\ee
In this case there are no single centered black holes carrying these charges
and the only two centered configurations which can contribute to
the index  carry charges of the form:
\be \label{eonly}
(a  Q, c Q)  \quad \hbox{and} \quad
(b P, d  P)\, ,
\ee
for some vectors $ Q$ and $ P$ and 
$\pmatrix{a & b\cr c & d}\in SL(2,\ZZZ)$, carrying total charge
$(a Q+b P,
c Q+d P)=(\wh Q, \wh P)$. 
This two centered configuration exists in a certain region of the moduli
space of the theory determined by the rules given in \cite{1104.1498}. Outside
this region the configuration ceases to exist and hence does not
contribute to the index.
The contribution to the index from this 
configuration when it exists is given by
\be \label{eeta}
(-1)^{ Q\cdot  P+1} | Q\cdot  P| \, d_h( Q^2 / 2) \, d_h ( P^2 / 2)\, ,
\ee
where
\be \label{edefdh}
\sum_n d_h(n) q^n = q^{-1} \prod_{k=1}^\infty 
\left(1 - q^k\right)^{-24}\, .
\ee
Physically $d_h(n)$ denotes the index of half BPS states.

The phenomenon of metamorphosis takes place when either $ P^2$
or $ Q^2$ (or both) take the value $-2$. Let us suppose $ P^2=-2$.
In that case the configuration
\be \label{eonly2}
(a'  Q',  c' Q')  \quad \hbox{and} \quad
(b' P' , d'  P')\, , \quad \pmatrix{a' & b'\cr c' & d'}
\equiv\pmatrix{a & b- au \cr c & d-cu}, \quad  Q'\equiv Q + u  P, \quad
 P'\equiv  P, \quad u \equiv  Q\cdot  P 
\ee
has the same total charge, satisfies 
\be \label{epqpq}
 Q^{\prime 2} =  Q^2, \quad 
 P^{\prime 2} =  P^2, \quad  Q'\cdot  P' =- Q\cdot  P\, ,
\ee
and hence, according to \refb{eeta} gives the same contribution to the index.
Now suppose that the configuration \refb{eonly} exists in the region $R_1$
in the moduli space and the configuration \refb{eonly2} exists in the region
$R_2$. It turns out that $R_1\cup R_2$ spans the whole moduli space of
the theory. Thus naively one would expect that in the region 
$R=R_1\cap R_2$ 
the
total contribution to the index from these two configurations
will be given by twice of \refb{eeta} whereas outside this region the
contribution to the index will be given by \refb{eeta}. {\it
However in order to match
the microscopic result we have to assume that in the region $R$ the
contribution to the index is given by \refb{eeta} while outside this region
there is no contribution to the index from these configurations.}

\begin{figure}

\begin{center}

\def\JPicScale{0.8}
\ifx\JPicScale\undefined\def\JPicScale{1}\fi
\unitlength \JPicScale mm
\begin{picture}(155,80)(0,0)
\linethickness{0.3mm}
\put(5,40){\line(1,0){35}}
\linethickness{0.5mm}
\multiput(20,40)(0.12,0.32){125}{\line(0,1){0.32}}
\linethickness{0.3mm}
\put(65,40){\line(1,0){35}}
\linethickness{0.5mm}
\multiput(115,80)(0.12,-0.95){42}{\line(0,-1){0.95}}
\linethickness{0.5mm}
\multiput(140,40)(0.12,0.32){125}{\line(0,1){0.32}}
\linethickness{0.3mm}
\put(115,40){\line(1,0){35}}
\put(20,35){\makebox(0,0)[cc]{0}}

\linethickness{0.5mm}
\multiput(65,80)(0.12,-0.95){42}{\line(0,-1){0.95}}
\put(65,35){\makebox(0,0)[cc]{$-u$}}

\put(120,35){\makebox(0,0)[cc]{$-u$}}

\put(140,35){\makebox(0,0)[cc]{0}}

\put(15,65){\makebox(0,0)[cc]{$R_1$}}

\put(75,65){\makebox(0,0)[cc]{$R_2$}}

\put(35,85){\makebox(0,0)[cc]{$L_1$}}

\put(65,85){\makebox(0,0)[cc]{$L_2$}}

\put(115,85){\makebox(0,0)[cc]{$L_2$}}

\put(155,85){\makebox(0,0)[cc]{$L_1$}}

\put(134,69){\makebox(0,0)[cc]{$R\equiv R_1\cap R_2$}}

\put(130,30){\makebox(0,0)[cc]{(c)}}

\put(25,30){\makebox(0,0)[cc]{(a)}}

\put(80,30){\makebox(0,0)[cc]{(b)}}

\end{picture}

\end{center}

\vskip -1in

\caption{Figure illustrating the walls of marginal stability and the region of
existence of the configurations described in \refb{eonly3} and \refb{eonly4}.
In Fig.~(a) the thick line $L_1$ 
labels the wall of marginal stability for
the configuration \refb{eonly3} which exists in 
the region $R_1$ to the left of $L_1$
in the upper half plane. In Fig.~(b) the thick line 
$L_2$ labels the wall 
of marginal stability for
the configuration \refb{eonly4} which exists in the region $R_2$ 
to the right of $L_2$
in the upper half plane. Fig.~(c) labels the region $R\equiv R_1\cap R_2$.
Thus naively we expect both configurations to exist in the region $R$ and one
of the two configurations to exist in the region outside $R$. However microscopic
counting requires that only one of the two configurations exist in the region
$R$ and none exist outside this region.
In drawing these figures we have implicitly taken $u$ to be positive. If $u$ is
negative then each figure has to be reflected about the vertical axis passing
through the origin.
}
\label{f1}
\end{figure}
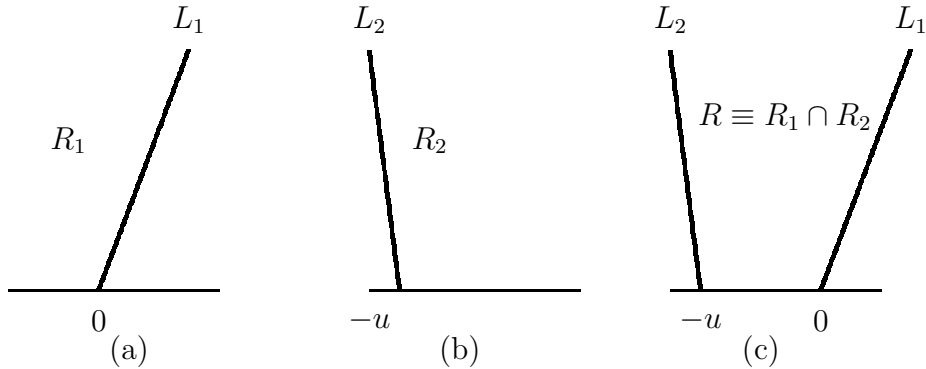

The case where $ Q^2=-2$ is related to the above by a duality transformation
and need not be discussed separately. In fact with the help of an 
S-duality transformation by the matrix $\pmatrix{d & -b\cr -c & a}$
we can map the configurations \refb{eonly} and \refb{eonly2} to 
\be \label{eonly3}
(Q, 0) \quad \hbox{and} \quad (0,P)\, , \quad P^2 =-2,
\ee
and
\be \label{eonly4}
(Q+uP, 0) \quad \hbox{and} \quad (-uP,P), \quad  u\equiv Q\cdot P\, ,
\ee
with each configuration carrying the same
index as \refb{eeta}.
Thus we shall focus on this configuration from now on. In this case Fig.~\ref{f1}
shows
the regions $R_1$, $R_2$ and $R$ in the upper half $\tau$-plane\cite{0702141} 
-- where
$\tau=\tau_1+i\tau_2$ denotes the asymptotic values of the axion-dilaton
modulus of the heterotic string theory on $T^6$-- for fixed asymptotic values of the
other fields. The boundaries of $R_1$, $R_2$ marked by the thick lines
$L_1$ and $L_2$ correspond to walls of marginal stability beyond which
the configurations \refb{eonly3} and \refb{eonly4} cease to exist. The precise
slope of these straight lines depend on the details of the charges and the
asymptotic values of the other moduli, and will be given in eqs.\refb{en2.23a} and
\refb{en2.29} respectively.

\begin{figure}

\begin{center}
\def\JPicScale{0.8}
\ifx\JPicScale\undefined\def\JPicScale{1}\fi
\unitlength \JPicScale mm
\begin{picture}(115,85)(0,0)
\linethickness{0.5mm}
\multiput(25,80)(0.12,-0.95){42}{\line(0,-1){0.95}}
\linethickness{0.5mm}
\multiput(70,40)(0.12,0.32){125}{\line(0,1){0.32}}
\linethickness{0.3mm}
\put(15,40){\line(1,0){100}}
\put(30,35){\makebox(0,0)[cc]{$-u$}}

\put(70,35){\makebox(0,0)[cc]{0}}

\linethickness{0.5mm}
\multiput(50,40)(0.24,1.95){21}{\multiput(0,0)(0.12,0.98){1}{\line(0,1){0.98}}}
\put(65,60){\makebox(0,0)[cc]{$R_1'$}}

\put(35,60){\makebox(0,0)[cc]{$R_2'$}}

\put(55,85){\makebox(0,0)[cc]{$L$}}

\put(25,85){\makebox(0,0)[cc]{$L_2$}}

\put(85,85){\makebox(0,0)[cc]{$L_1$}}

\end{picture}
\end{center}

\vskip -1in

\caption{The pictorial description of black hole metamorphosis.}
\label{f2}
\end{figure}
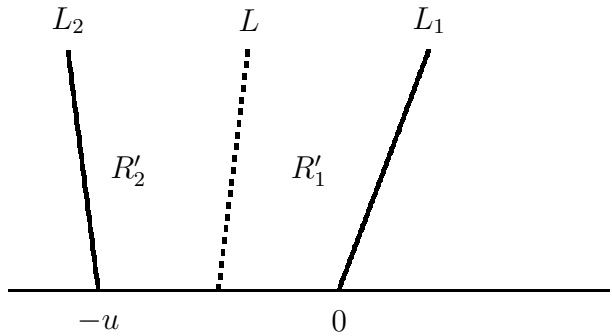

The phenomenon of black hole metamorphosis suggests the existence
of a hypothetical line $L$, shown in Fig.~\ref{f2}, 
such that the configuration \refb{eonly3} exists
only in the region $R_1'$ 
to the right of $L$ and left of $L_1$
and the configuration \refb{eonly4} exists only in the region $R_2'$
to the left of $L$ 
and
the right of $L_2$. In that case it would explain why
the index is given by \refb{eeta} in the region $R_1'\cup R_2'=R$ and vanishes
outside this region. Our goal will be to understand the physical origin of $L$.

\sectiono{Review of multi-black hole solutions in $N=2$ supergravity} \label{s1.5}

Although heterotic string theory on $T^6$ describes an $\NN=4$ supersymmetric
string theory,  the multi-black hole solutions are best understood 
 in the language of $\NN=2$ supergravity. For this reason in this
section we shall review multi-black hole solutions in $\NN=2$ supergravity.
The bosonic 
fields of an $\NN=2$ supergravity coupled to $n_v$ vector multiplet fields 
are the metric $g_{\mu\nu}$, $n_v+1$ complex scalars $X^I$,
and $n_v+1$ gauge fields $\AAA_\mu^I$ with $0\le I\le n_v$. 
The theory has a complex gauge invariance under which all the $X^I$'s scale by an
arbitrary complex function $\Lambda(x)$, the metric scales by $|\Lambda|^{-2}$
and the gauge fields remain invariant. 
The action of the theory is completely fixed by the prepotential
$F$ which is a meromorphic, homogeneous function of the $X^I$'s of degree 2. 
If $(q,p)$ denote the electric and
magnetic charge vectors carried by a state with $q$ and $p$ being $n_v+1$
dimensional vectors, then we define
\be \label{en2.1}
F_I\equiv \p F/ \p X^I, \quad e^{-K}\equiv i(\bar X^I F_I - X^I \bar F_I),
\quad Z(q,p) \equiv ( q_I X^I-p^I F_I ) e^{K/2}\, .
\ee
The gauge fields are normalized so that the action of
a test particle carrying electric charges $\hat q_I$ and magnetic charges $\hat p^I$
takes the form
\be \label{ed1}
{1\over 2} \int ( \hat q_{I} \AAA_\mu^I- \hat p^I \AAA_{I\mu}) {dx^\mu}
\ee
where $\AAA_\mu^I$ denotes the usual gauge potential and $\AAA_{I\mu}$
denotes the dual magnetic potential. 

A general supersymmetric multi-centered black hole solution in such a theory was
constructed in \cite{0005049,0304094}. To describe the solution we introduce the functions:
\be \label{en2.4}
H^I = \sum_i {p^I_{(i)}\over |\vec r -\vec r_i|} - 2 \, {\rm Im} \left[ e^{-i\alpha_\infty} X^I
e^{K/2}\right]_\infty,
\quad 
H_I = \sum_i {q_{(i)I}\over |\vec r -\vec r_i|} - 2 \, {\rm Im} \left[ e^{-i\alpha_\infty} F_I
e^{K/2}\right]_\infty\, ,
\ee
where $\vec r_i$ are the locations of the centers in the three dimensional space,
$(q_{(i)}, p_{(i)})$ denote the electric and magnetic charges carried by the $i$-th center,
the subscript $~_\infty$ denotes the asymptotic values of the various fields and
\be \label{en2.5}
\alpha_\infty = {\rm Arg}\left[Z\left(\sum_i q_{(i)}, \sum_i p_{(i)}\right)\right]_\infty\, .
\ee
Now
let
\be \label{en2.16}
S_{BH}(\{q_I\}, \{p^I\}) = \pi\, \Sigma(\{q_I\}, \{p^I\})\, ,
\ee
be the entropy of a single centered black hole solution in this theory with charge $(q,p)$.
There is a standard algorithm for computing the function $\Sigma$ from the 
knowledge
of the function $F$ -- it is given by the extremum of $|Z(q,p)|^2$ with respect
to the scalar moduli fields.
We now define
\be \label{en2.16.1}
\chi^K(\{q_I\}, \{p^I\}) \equiv  -{\p\Sigma \over \p q_K}, \quad
\chi_K(\{q_I\}, \{p^I\}) \equiv {\p\Sigma \over \p p^K}\, ,
\ee
\be \label{en2.17} 
g^K(\{q_I\}, \{p^I\}) = \chi^K(\{q_I\}, \{p^I\}) + i p^K,
\quad 
g_K(\{q_I\}, \{p^I\}) = \chi_K(\{q_I\}, \{p^I\})+ i q_K\, .
\ee
Then the solution for the scalar fields, metric and the gauge fields is given by
\be \label{en2.18}
{X^K\over X^0} = {g^K(\{H_I(\vec r)\}, \{H^I(\vec r)\})\over
g^0(\{H_I(\vec r)\}, \{H^I(\vec r)\})}, \quad
{F_K\over X^0} = {g_K(\{H_I(\vec r)\}, \{H^I(\vec r)\})\over
g^0(\{H_I(\vec r)\}, \{H^I(\vec r)\})}\, ,
\ee
\be \label{en2.19}
ds^2 = e^{2 V} (dt + \vec \omega\cdot d\vec {x})^2 + e^{-2V} dx^i dx^i \, ,
\ee
\be \label{en2.20}
e^{-2V} \equiv \Sigma(\{H_I(\vec r)\}, \{H^I(\vec r)\})\, ,
\ee
\ben \label{en2.20.3}
\AAA^K_\mu dx^\mu &=& -\Sigma(\{H_I(\vec r)\}, \{H^I(\vec r)\})^{-1}
\chi^K(\{H_I(\vec r)\}, \{H^I(\vec r)\})
(dt + \vec \omega\cdot d\vec {x}) - \sum_{i} p_{(i)}^K
\cos\theta_{(i)} d\phi_{(i)}, \nonumber \\
\AAA_{K\mu} dx^\mu &=& -\Sigma(\{H_I(\vec r)\}, \{H^I(\vec r)\})^{-1}
\chi_K(\{H_I(\vec r)\}, \{H^I(\vec r)\}) 
(dt +\vec \omega\cdot d\vec {x})- \sum_{i} q_{(i)K}
\cos\theta_{(i)} d\phi_{(i)} \nonumber \\
\een
where $(\theta_{(i)}, \phi_{(i)})$ denote the polar and azimuthal angles of the
spherical polar coordinate system with origin at $\vec r_i$.
The general solution for $\vec\omega$ exists but we shall not need it. For
single centered solution $\vec\omega$ vanishes.
Finally consistency demands that the locations $\vec r_i$ be subject to the
constraint:
\be \label{ed2}
\sum_{j=1\atop j\ne i}^n { q_{(i)I} p^I_{(j)}- q_{(j)I} p^I_{(i)}\over |\vec r_i - \vec r_j|}
= 2 \, {\rm Im} \, (e^{-i\alpha_\infty} Z_i), \quad Z_i \equiv Z\left.\left(q_{(i)}, p_{(i)}\right)
\right|_{\infty}
\ee

One clarification in necessary here. The combinations $X^I/X^0$, $F_I/X^0$ and the
gauge fields are invariant under the complex gauge transformation generated by
$\Lambda(x)$ and hence it is not necessary to specify in which gauge we have given the
solutions. However since the metric is not invariant under this transformation we need to
specify the gauge in which the metric is given. \refb{en2.19} is given in the choice of gauge
in which the Einstein-Hilbert term takes the form\cite{0304094}
\be \label{ehilbert}
{1\over 16\pi} \int d^4 x \, \sqrt{-\det g} \, R\, .
\ee

For a 2-centered solution carrying charges $(q_{(1)}, p_{(1)})$ at $\vec r_1$ and
$(q_{(2)}, p_{(2)})$ at $\vec r_2$, \refb{ed2} gives
\be \label{en2.20.1}
|\vec r_1-\vec r_2| = { q_{(2)I} p^I_{(1)}- q_{(1)I} p^I_{(2)}\over 2 \, {\rm Im} 
(e^{-i\alpha_\infty} Z_2)}, \quad e^{i\alpha_\infty} ={Z_1+Z_2
\over |Z_1+Z_2|}\, .
\ee
When $|Z(q_{(2)}, p_{(2)})|$ is small and we can ignore the background field
produced by the second center in most of the space, then 
an independent way of arriving at the result \refb{en2.20.1} is as follows. Let us
consider the background fields produced by a single centered solution carrying charges
$(q_{(1)}, p_{(1)})$ placed at the origin. If we now place a test particle
carrying charge $(\hat q, \hat p)$ in this background then the 
action of this test particle takes the form
\be \label{en2.20.2}
S_t =  - \int d\tau \, |Z(\{\hat q_I\}, \{\hat p^I\})| 
 + {1\over 2} \int (\hat q_{I} \AAA_\mu^I - \hat p^I \AAA_{I\mu}) {dx^\mu}
\ee
where $\tau$ is the proper time, and $x^\mu$ denote the coordinates of the test
particle. 
If the test particle is at rest then we have $d\tau = e^{V} dt$ and hence
\be \label{en2.20.2.1}
S_t =   \int dt \, \left[-e^V\, |Z(\{\hat q_I\}, \{\hat p^I\})| 
 + {1\over 2} ( \hat q_{I} \AAA_0^I-\hat p^I \AAA_{I0})\right]\, .
\ee
The equilibrium position of the test particle will be at the extremum
of the integrand with respect to the spatial coordinates 
$x^1,x^2,x^3$. 
It can be shown that this gives us back \refb{en2.20.1}
with $(q_{(2)}, p_{(2)})$ replaced by $(\hat q, \hat p)$ if $|Z(q_{(2)}, p_{(2)})|$
is small so that we can treat the second center as a test particle
ignoring its backreaction on the geometry\cite{0005049}.

\sectiono{S-T-U model} \label{s1}

In this section we shall analyze a class of 2-centered black hole solutions in heterotic
string theory on $T^6$ and propose a mechanism for black hole metamorphosis.
Our analysis will proceed in several steps. First we shall describe a truncation of heterotic
string theory on $T^6$ which can be mapped to an $\NN=2$ supergravity theory, known as the
S-T-U model. We shall
then describe the S-T-U model and the maps between the fields in the  
two descriptions. We then consider a two centered configuration in this theory with one
center carrying charge $(0,P)$ with $P^2=-2$, 
and take a limit where the other center carrying charge
$(Q,0)$ becomes light and can be regarded as
a probe. The technique reviewed in \S\ref{s1.5} 
then enables us to easily construct the background
field associated with the heavy center and find 
the equilibrium position of the light center.
We then analyze the solution carefully to find the region of the moduli space where it
exists. Although naively it exists in the region $R_1$ to the left of the line
$L_1$ in Fig.~\ref{f2}, we 
suggest a mechanism by which the region of existence gets truncated to 
$R_1'$ displayed in Fig.~\ref{f2}. We repeat the analysis for another configuration related to
the first by the transformation \refb{eonly4} and show that this exists in the region $R_2'$ as
displayed in Fig.~\ref{f2}. This analysis also allows us to determine the precise location
of the line $L$ in Fig.~\ref{f2}.

\subsection{Truncation of heterotic string theory on $T^6$} \label{strunc}

We shall now describe the truncation of heterotic string theory on $T^6$ that can be
mapped to an $\NN=2$ supergravity theory. 
For this we take $T^6$ in the form of the
product $T^4\times T^2$ and ignore all excitations of the components of the 
metric and 2-form fields with one or both legs along $T^4$ and also all 
excitations of the ten dimensional gauge fields. This truncated theory will have only
four gauge fields corresponding to 4-$\mu$ and 5-$\mu$ components of the metric
and the 2-form gauge fields, with $x^4$ and $x^5$ denoting the 
coordinates along $T^2$ and $x^\mu$ with $0\le \mu\le 3$ denoting the coordinates
along the 3+1 dimensional non-compact space-time.
The other relevant fields are the canonical metric $g_{\mu\nu}$, the axion dilaton
modulus $S=S_1+iS_2$, the complex structure modulus $U=U_1+i U_2$ 
of $T^2$ and the complexified
Kahler modulus $T=T_1 + i T_2$ of $T^2$. 
The four components $(Q_1, Q_2, Q_3, Q_4)$ of the electric charge
vector $Q$ correspond respectively to the number of units of momentum
along $x^5$ and $x^4$ respectively and winding numbers  along $x^5$ and
$x^4$ respectively. On the other hand the components $(P_1,P_2, P_3,P_4)$
of the magnetic charge $P$ denote respectively the heterotic five-brane winding
numbers along $T^4\times x^4$-circle and $T^4\times x^5$-circle 
respectively and Kaluza-Klein
monopole charges associated with $x^5$ and $x^4$ directions
respectively. The bilinears $Q^2$,
$P^2$, $Q\cdot P$ are given by
\be \label{ebi1}
Q^2 = 2(Q_1 Q_3 + Q_2 Q_4), \quad P^2 = 2(P_1 P_3 + P_2 P_4), \quad
Q\cdot P= Q_1 P_3 + Q_2 P_4 + Q_3 P_1 + Q_4 P_2 \, .
\ee
Finally the entropy of a black hole carrying (electric, magnetic) charges $(Q,P)$
is given by
\be \label{ebi2}
S_{BH} = \pi \sqrt\Sigma, \quad \Sigma ={Q^2 P^2 - (Q\cdot P)^2}\, .
\ee

\subsection{$\NN=2$ description} 

This truncated theory can be mapped to an $\NN=2$
supergravity theory coupled to three vector multiplets, with prepotential
\be \label{en2.2}
F = -{X^1 X^2 X^3\over X^0}\, .
\ee
The scalar fields $S$, $T$ and $U$ introduced in \S\ref{strunc} are given by
\be \label{en2.2.1}
S = {X^1\over X^0}, \quad 
T = {X^2\over X^0} , \quad
U={X^3\over X^0} \, .
\ee
The relations between the gauge fields in the two descriptions can be described
by giving the relations between the charges $\{Q_i, P_i\}$ given above with the
charges $\{ q_I, p^I\}$ in the $\NN=2$ supergravity description. This is as 
follows (see {\it e.g.} \cite{0708.1270} for a review)
\be \label{en2.6}
Q\equiv (Q_1,Q_2,Q_3,Q_4)=(q_0,q_3, -p^1,q_2), \quad
P\equiv (P_1,P_2,P_3,P_4)=( q_1,p^2,p^0,p^3)\, .
\ee
Eq.\refb{ebi2} now gives
\be \label{en2.15}
\Sigma(\{q_I\}, \{p^I\}) = \left[4 (q_2 q_3 - q_0 p^1) (p^0 q_1 + p^2 p^3) 
- (q_0 p^0 - q_1 p^1 + q_2 p^2 + q_3 p^3)^2\right]^{1/2}\, .
\ee
We shall denote the asymptotic values of the various moduli fields as 
\be \label{en2.3}
S|_\infty =\zeta \equiv \zeta_1+i\zeta_2, \quad 
T|_\infty= \rho
\equiv\rho_1+i\rho_2, 
\quad U|_\infty =\sigma \equiv \sigma_1+i\sigma_2\, .
\ee
As we shall see in \refb{ezetatau}, 
$\zeta$ is related to the modulus $\tau$ of \S\ref{smeta}
via the relation $\zeta =-\bar\tau$.
We also define
\be \label{en2.3.1}
x^0 \equiv X^0_\infty\, .
\ee
{}From \refb{en2.1}, \refb{en2.2},  \refb{en2.3} and \refb{en2.3.1} 
it follows that 
\be \label{en2.3.5}
e^{-K} = 8 X^0 \bar X^0 S_2 T_2 U_2, \quad 
\, \left. e^{-K}\right|_\infty=8\, x^0 \bar x^0 \zeta_2\sigma_2 \rho_2\, .
\ee

\subsection{The two centered solution} \label{s2c}

In the asymptotic background described above
we construct a two centered solution with the
first center carrying charge $(0,P)$ and the second center carrying charge 
$(Q,0)$, with
\be \label{en2.7}
Q=(a,b,c,d), \quad P=(0,-1,0,1)\, .
\ee
This gives, from \refb{ebi1}
\be \label{en2.8}
Q^2 = 2ac+2bd, \quad P^2 = -2, \quad u\equiv Q\cdot P=b-d\, .
\ee
We shall for definiteness take $(b-d)>0$ so that $u>0$. 
Since $P^2=-2$ this configuration should display the phenomenon of black hole bound
state metamorphosis. In particular there must exist a line $L$ in the $\tau=
-\bar\zeta$ plane
such that  the bound state ceases to exist to the left of this line. Our goal will be
understand the physical origin of this hypothetical line $L$.

Now using \refb{en2.6} we see that in the language of $\NN=2$ supergravity the two
centers carry charges $(q_{(1)}, p_{(1)})$ and $(q_{(2)}, p_{(2)})$ where
\be \label{en2.9}
p_{(1)} = (0, 0, -1, 1), \quad q_{(1)}=(0,0,0,0), \qquad
p_{(2)} = (0, -c, 0, 0), \quad q_{(2)}=(a, 0, d, b)\, .
\ee
We define
\ben \label{en2.10}
&& Z_1\equiv Z(q_{(1)}, p_{(1)})|_\infty = \left[ e^{K/2}(F_2-F_3)\right]_\infty = 
\sqrt{x^0\over \bar x^0}
{1\over \sqrt{8\zeta_2\rho_2\sigma_2}} \, \zeta(\rho-\sigma)\, ,
\nonumber \\
&& Z_2  \equiv Z(q_{(2)}, p_{(2)}) |_\infty=  \left[ e^{K/2}(aX^0 + d X^2
+ b X^3 +cF_1)\right]_\infty \nonumber \\
&& \qquad \qquad \qquad \qquad \,  \, \, \, = \sqrt{x^0\over \bar x^0}
{1\over \sqrt{8\zeta_2\rho_2\sigma_2}} \, (a + d\rho+ b\sigma - c\rho\sigma)
\, . 
\een
To simplify the analysis we shall work in the limit where $\zeta_2$ is large.
In this limit $|Z_2|$ given in \refb{en2.10} is small showing that the corresponding state
is light. Hence we can ignore its effect on the background field and treat this center
as a probe. In this limit the background geometry approaches that of a single
centered black hole with charge $(q_{(1)}, p_{(1)})$ placed at $\vec r_1$,
and $\alpha_\infty$ defined in 
\refb{en2.5} and 
the functions $H^I$ and $H_I$ introduced in \refb{en2.4} take the form
\be \label{en2.11}
e^{i\alpha_\infty} = {Z_1+Z_2 \over  |Z_1+Z_2|}\simeq  {Z_1 \over  |Z_1|} = 
\sqrt{x^0\over \bar x^0}
{\zeta \over |\zeta |}
{\rho -\sigma  \over |\rho -\sigma |}\, ,
\ee
\be \label{en2.12}
(H^0, H^1, H^2, H^3) \simeq 
{1\over |\vec r -\vec r_1|} (0,0,-1, 1) -  {2\over \sqrt{8\zeta_2\rho_2\sigma_2}} 
{\rm Im} \left\{
{|\zeta |\over \zeta }\, {| \rho-\sigma  |\over\rho- \sigma }
\, 
(1, \zeta , \rho , \sigma ) \right\} \, ,
\ee
and
\be  \label{en2.13}
(H_0, H_1, H_2, H_3) \simeq 
{2\over \sqrt{8\zeta_2\rho_2\sigma_2}} 
{\rm Im} \left\{
{|\zeta |\over \zeta }\, {|\sigma  - \rho |\over\rho- \sigma  }
\, 
(-\zeta  \rho  \sigma , \rho  \sigma , \zeta  \sigma , 
\zeta  \rho  ) \right\} \, .
\ee
{}From this we can construct the solution for the metric, scalars and gauge fields
using the prescription reviewed in \S\ref{s1.5}. 
The separation between the two centers is given, according to \refb{en2.20.1},  by 
\be \label{en2.14}
|\vec r_1-\vec r_2| =  {b-d\over 2} \, {\sqrt{8 \zeta_2 \sigma_2 \rho_2}\over |\zeta| |\sigma-\rho|}
{1 \over {\rm Im} \left[ (a + d\rho + b\sigma - c\rho\sigma) / (\zeta (\rho-\sigma))
\right]}\, .
\ee

Before we go on we must mention two subtle points in the relation between the 
$\NN=4$ and
$\NN=2$ theory that will be important for our analysis. 
According to \refb{en2.10}, the total
mass of the system is given by
\ben \label{etotal1}
|Z_1+Z_2| &=& {1\over \sqrt{8\zeta_2\rho_2\sigma_2}} \, \sqrt{|A|^2 + |B|^2 |\zeta|^2 
+ 2\, \zeta_1 {\rm Re} \, (AB^*) + 2\, \zeta_2 \, {\rm Im}\, (AB^*)}, \nonumber \\ && \quad
A \equiv a + d\rho+ b\sigma - c\rho\sigma, \quad B\equiv 
(\rho-\sigma)\, .
\een
Now consider a state carrying total charge $(P,P)$. The BPS mass of this state will
be given by  setting $a=c=0$ and $b=-1, d=1$ in \refb{etotal1}
and its dependence on the axion dilaton modulus $\zeta$ will be proportional to
$\sqrt{|1+\zeta|^2} /\sqrt{\zeta_2}$. On the other hand in the convention of
\cite{0702141,1104.1498} which we used in presenting the results in
\S\ref{smeta}, the dependence of the BPS mass of a particle of charge $(P,P)$
on the axion dilaton   modulus is proportional to $\sqrt{|1-\tau|^2} /\sqrt{\tau_2}$.
This shows that $\zeta$ and $\tau$ are related by
\be \label{ezetatau}
\zeta=-\bar\tau\, .
\ee
To discuss the second subtlety, 
let us  return to the general formula \refb{etotal1}.
The BPS mass formula in the $\NN=4$ supersymmetric
theories (derived in \cite{9507090,9508094} and
used {\it e.g.} in \cite{0702141} for the analysis of the walls of marginal stability)
is given by the same formula as \refb{etotal1} (after the identification
\refb{ezetatau}) except that the coefficient of $\zeta_2=\tau_2$
under the square root is given by $2|{\rm Im}\, (AB^*)|$. Thus the two formul\ae\ agree
when ${\rm Im}\, (AB^*)>0$, \i.e.\
\be \label{etotal2}
(\sigma_2-\rho_2) (a + d\rho_1 + b\sigma_1 - c\rho_1\sigma_1+
c\rho_2\sigma_2) +(\rho_1 - \sigma_1) (d\rho_2 + b\sigma_2 - c(\rho_2 \sigma_1+\rho_1\sigma_2)) 
>0\, .
\ee
{}From now on we shall assume that this condition holds.

\subsection{The region of existence of the solution} 

{}From \refb{en2.14} we can identify the wall of marginal stability as the curve in
the $\zeta$ plane on which the right hand side of \refb{en2.14} diverges. This gives
\be \label{en2.21}
{\zeta_1 \over \zeta_2} = {N\over D}, \quad \i.e. \quad {\tau_1\over \tau_2}=-{N\over D}
\, ,
\ee
where 
\ben \label{en2.22}
N &=& -(\sigma_2-\rho_2) (d\rho_2 + b\sigma_2 - c(\rho_2 \sigma_1+\rho_1\sigma_2))
+(\rho_1 - \sigma_1) (a + d\rho_1 + b\sigma_1 - c\rho_1\sigma_1+
c\rho_2\sigma_2)\, ,\nonumber \\
D &=& (\sigma_2-\rho_2) (a + d\rho_1 + b\sigma_1 - c\rho_1\sigma_1+
c\rho_2\sigma_2) +(\rho_1 - \sigma_1) (d\rho_2 + b\sigma_2 - c(\rho_2 \sigma_1+\rho_1\sigma_2)) 
\, . \nonumber \\
\een
\refb{en2.21} marks the location of the line $L_1$ in Figs.~\ref{f1} and \ref{f2}.
In particular the solution exists when the right hand side of \refb{en2.14} is positive, \i.e.\
for
\ben \label{en2.23}
{\zeta_1} & > &{N\over D} \zeta_2\quad \hbox{for $(b-d)D>0$}\, ,\nonumber \\
&  < &{N\over D} \zeta_2 \quad \hbox{for $(b-d)D<0$}\, .
\een
Since according to \refb{etotal2} we have $D>0$, and we have assumed that
$b-d>0$, this condition translates to
\be \label{en2.23a}
{\zeta_1}   > {N\over D} \zeta_2 \quad \i.e. \quad {\tau_1 } < -{N\over D}\,
\tau_2\, .
\ee

Naively one may expect that \refb{en2.23a} is the only condition
on $\tau$ for the existence of the solution, since as long as \refb{en2.23a}
is satisfied, $|\vec r_1-\vec r_2|$ given in \refb{en2.14} remains positive. 
However upon
closer examination one discovers a peculiar property of the solution that can be
attributed to the special charge vector carried by the first center. If we take a
test particle of charge $(P,0)$ with $P=( 0, -1, 0, 1)$ as in \refb{en2.7}, it maps to
$q=(0,0,1,-1)$, $p=(0,0,0,0)$ in the $\NN=2$ supergravity variables, and its
mass at a point $\vec r$ is given by
\be \label{ecentral}
{1\over 8\sqrt{S_2(\vec r)T_2(\vec r)U_2(\vec r)}} | T(\vec r) - U(\vec r)| \, .
\ee
Thus it vanishes when $T(\vec r)=U(\vec r)$. Using 
eqs.\refb{en2.16.1}-\refb{en2.18} and \refb{en2.15} we see that this requires
$H_2=H_3$ and $H^2=H^3$. Now from \refb{en2.13} we see that the first
condition is satisfied automatically, while eq.\refb{en2.12} tells us that we
have $H^2=H^3$ when
\be \label{en2.24}
|\vec r - \vec r_1| 
= r_e, \quad r_e 
\equiv \sqrt{8\zeta_2 \sigma_2 \rho_2} \, {|\zeta|\over \zeta_2 |\rho-\sigma|}\, .
\ee
This describes a spherical shell of radius $r_e$
around $\vec r_1$ on which an electrically charged test
particle carrying charge $(P,0)$ becomes
massless. Physically on this shell the radius of the $x^4$ direction
reaches the self-dual point and hence we have massless non-abelian
gauge fields. This in turn shows that at this point the original 
solution describing the background field produced by the charge
$(0,P)$ breaks down and we should not trust the solution 
for values of $\vec r$ for which $|\vec r - \vec r_1|$ 
is less than $r_e$ defined in  \refb{en2.24}.
This has been named the enhancon mechanism in \cite{9911161}. 
Indeed, if we ignore this effect and continue to trust the solution
for $|\vec r - \vec r_1| < r_e$, then at some point
$\Sigma(\{H_I\}, \{H^I\})$ computed from \refb{en2.15},
\refb{en2.12}  and \refb{en2.13} vanishes and
the solution becomes singular\cite{9911161}.
We shall call $r_e$ the enhancon radius.
Thus a two centered solution, obtained by placing in the above background
a test charge $(Q,0)$ at $\vec r_2$ is sensible only when we have
\be \label{exists}
|\vec r_1 - \vec r_2| 
> r_e\, .
\ee
Using \refb{en2.14}, \refb{en2.24}
and the positivity of $D$ and $b-d$, this translates to
\be \label{en2.25}
{\zeta_1} < {b-d\over 2D} \zeta_2 |\rho-\sigma|^2 + {N\over D}\zeta_2
\quad \i.e. \quad 
{\tau_1} > -{b-d\over 2D} \tau_2 |\rho-\sigma|^2 - {N\over D}\tau_2\, .
\ee

As we shall see in \S\ref{s2.2},
the correct description of the solution involves replacing it
by a gravitationally dressed smooth BPS dyon obtained by boosting the
Harvey-Liu monopole 
solution\cite{harliu,0110109} in an internal compact direction. As a result
the solution
begins to differ from that given in \S\ref{s2c} even for
$|\vec r - \vec r_1|> r_e$. However for now we shall 
take the above bound on $\zeta_1$ seriously and examine its consequences.
In this case
\refb{en2.25} gives us the location of
the left boundary $L$ of the region $R_1'$ in Fig.~\ref{f2}, with the right 
boundary $L_1$ of $R_1'$ being given
by the wall of marginal stability \refb{en2.23a}. In \S\ref{s2.2} we
shall see that this in fact is the exact result for the allowed range of $\zeta_1$
in the large $\zeta_2$ limit.

\subsection{The second two centered solution}

Next consider the two centered configuration with charges
$(-uP, P)$ and $(Q+uP, 0)$ where $u=Q\cdot P=(b-d)$. 
Again one can argue that in the limit of large $\zeta_2$ the second center
carrying only electric charge $(Q+uP,0)$ is light and hence can be
treated as a test particle. Furthermore, for the first center the contribution
to the background field from the electric component proportional to
$uP$ will be small and hence can be dropped. Thus the problem effectively
reduces to studying the test charge $(Q+uP,0)$ in the background produced
by the charge $(0,P)$. 
Since according to \refb{en2.7}, \refb{en2.8}, 
$Q+uP$ differs from $Q$ just by the exchange of the quantum numbers $b$
and $d$, we can derive the various results for this system simply by exchanging $b$
and $d$ in the earlier results. In particular
for this system the separation between the two centers is given by
\be \label{en2.14aa}
|\vec r_1-\vec r_2| =  {d-b\over 2} \, {\sqrt{8 \zeta_2 \sigma_2 \rho_2}\over |\zeta| |\sigma-\rho|}
{1 \over {\rm Im} \left[ (a + b\rho + d\sigma - c\rho\sigma) / (\zeta (\rho-\sigma))
\right]}\, .
\ee
The wall of marginal
stability, where $|\vec r_1-\vec r_2|$ diverges, is at
\be \label{en2.27}
{\zeta_1} = {N'\over D'}  \zeta_2 \quad \i.e. \quad {\tau_1} = -{N'\over D'}  \tau_2\, ,
\ee
where
\ben \label{en2.28}
N' &=& -(\sigma_2-\rho_2) (b\rho_2 + d\sigma_2 - c(\rho_2 \sigma_1+\rho_1\sigma_2))
+(\rho_1 - \sigma_1) (a + b\rho_1 + d\sigma_1 - c\rho_1\sigma_1+
c\rho_2\sigma_2)\, , \nonumber \\
D' &=& (\sigma_2-\rho_2) (a + b\rho_1 + d\sigma_1 - c\rho_1\sigma_1+
c\rho_2\sigma_2) +(\rho_1 - \sigma_1) (b\rho_2 + d\sigma_2 - c(\rho_2 \sigma_1+\rho_1\sigma_2)) 
\nonumber \\
&=& D\, .
\een
\refb{en2.27} marks the location of the line $L_2$ in Figs.\ref{f1} and \ref{f2}.
The solution exists for
\be \label{en2.29}
{\zeta_1}  <  {N'\over D} \zeta_2 \quad \i.e. \quad {\tau_1}  >  -{N'\over D} \tau_2\, .
\ee
The enhancon radius remains at the same place as before. The condition
that the location of
the second center lies outside the enhancon radius can be translated to
\be \label{en2.30}
{\zeta_1} > {d-b\over 2D} \zeta_2 |\rho-\sigma|^2 + {N'\over D}\zeta_2
\quad \i.e. \quad 
{\tau_1} < {b-d\over 2D} \tau_2 |\rho-\sigma|^2 - {N'\over D}\tau_2\, .
\ee
As before we shall take this to be our estimate for the right boundary
of the region $R_2'$ in Fig.~\ref{f2}, with the left boundary $L_2$ 
of $R_2'$ being
given by the constraint \refb{en2.29}.

We now note that
\be \label{erel1}
{b-d\over 2D}  |\rho-\sigma|^2 - {N'\over D} = 
{d-b\over 2D}  |\rho-\sigma|^2 - {N\over D} \, ,
\ee
\i.e.\ the right hand sides of \refb{en2.25} and \refb{en2.30} match. This in turn
shows that the right boundary of $R_2'$ coincides with the left boundary $L$ of
$R_1'$, and hence in any region of the moduli space between the two walls
of marginal stability $L_1$ and $L_2$ in Fig.~\ref{f2}, one and only
one of the two configurations exists. This is precisely what is  required for the
black hole metamorphosis hypothesis to hold.

\subsection{Special case of diagonal $T^6$} \label{sdiag}

\newcommand{\sign}{\rm sign}

For later use we shall now write down the explicit solutions in the
special case
\be \label{en2.35}
\sigma_1=\rho_1=0\, ,
\ee
corresponding to setting the off-diagonal components of the metric
and the 2-form field along $T^2$ to zero at infinity.
Furthermore we shall take the location of the first center at the origin so that
\be \label{en2.35.1}
\vec r_1=0, \quad |\vec r - \vec r_1|=r\, .
\ee
Then we can express \refb{en2.12}, \refb{en2.13} as
\ben\label{en2.36}
H^0 &=& {2\over \sqrt{8\zeta_2\rho_2\sigma_2}} \, \sign(\rho_2-\sigma_2)
{\zeta_1\over |\zeta|} \, ,
\nonumber \\
H^1 &=& {2\over \sqrt{8\zeta_2\rho_2\sigma_2}} \, \sign(\rho_2-\sigma_2) |\zeta|\, ,
\nonumber \\
H^2 &=& -{1\over r}+ 
{2\over \sqrt{8\zeta_2\rho_2\sigma_2}} \, \sign(\rho_2-\sigma_2)
{\rho_2\zeta_2\over |\zeta|}\, ,
\nonumber \\
H^3 &=& {1\over r}+ 
{2\over \sqrt{8\zeta_2\rho_2\sigma_2}} \, \sign(\rho_2-\sigma_2)
{\sigma_2\zeta_2\over |\zeta|}\, ,
\nonumber \\
H_0 &=& -{2\over \sqrt{8\zeta_2\rho_2\sigma_2}} \, \sign(\rho_2-\sigma_2)
{\rho_2\sigma_2} |\zeta| \, ,
\nonumber \\
H_1 &=& {2\over \sqrt{8\zeta_2\rho_2\sigma_2}} \, \sign(\rho_2-\sigma_2)
{\rho_2\sigma_2\zeta_1\over |\zeta|} \, ,
\nonumber \\
H_2 &=& 0 \, ,
\nonumber \\
H_3 &=& 0\, .\een
This gives from \refb{en2.15}, \refb{en2.16.1}-\refb{en2.20.3}, 
\ben \label{en2.37}
&&\Sigma(\{H_I\}, \{H^I\}) = [-4 H_0 H^1 H^2 H^3]^{1/2}\, , \nonumber \\
&& S = {X^1\over X^0}=
{2 i H_0 H^1\over -\Sigma +2 i H_0 H^0}\, , \quad
T = {X^2\over X^0} = -{2i H_0 H^2 \over \Sigma}, \quad U = {X^3\over X^0}
= -{2 i H_0 H^3 \over \Sigma}\, , \nonumber \\
ds^2 &=& -\Sigma^{-1}\,  dt^2 + \Sigma\,  dx^i dx^i
\, , \nonumber \\
\AAA^{0}_\mu dx^\mu  &=&  -{2 H^1 H^2 H^3\over \Sigma^2} \, dt , \qquad
\AAA^2_\mu dx^\mu = -{2 H_0 H^0 H^2\over \Sigma^2}\, dt +\cos\theta d\phi, 
\nonumber \\
\AAA^3_\mu dx^\mu &=& -{2 H_0 H^0 H^3\over \Sigma^2}\, dt - \cos\theta d\phi, \qquad
\AAA_{1\mu} dx^\mu = {2 H_0 H^2 H^3\over \Sigma^2}\, dt
\, .
\een
Note that we have  given the expressions for the electric potentials 
$\AAA^{0}_\mu,\AAA^{2}_\mu,\AAA^{3}_\mu$ and the magnetic potential 
$\AAA_{1\mu}$. This contains full information about all the gauge fields.
Eq.\refb{en2.37} shows that $T$ and $U$ remain purely imaginary for
all values of $\vec r$ and hence the off diagonal components of the metric and
the 2-form field along $T^2$ continue to vanish at all points.

{}From eqs.\refb{en2.20.2.1} and \refb{en2.9} 
we get the Lagrangian of the test particle carrying charge
$(q_{(2)}, p_{(2)})$ in this background to be
\ben \label{en2.40}
L_t  &=& -\Sigma(\{H_I\}, \{H^I\})^{-1/2} {1\over \sqrt{8S_2(\vec r) T_2(\vec r)
 U_2(\vec r)}} |a + d \, T(\vec r) + b \, U(\vec r) - c\, T(\vec r) \, U(\vec r)|
\nonumber \\ && + {1\over 2} (c \, \AAA_{10} + a \, \AAA^0_0 + d \, \AAA^2_0 + b 
\, \AAA^3_0)\nonumber \\
&=& -{1\over 4 H_0} \left\{1 +{H^0 H_1\over H^2 H^3}\right\}^{1/2} 
\left[ \left( a - c{H_0\over H^1}\right)^2 -{H_0\over H^1 H^2 H^3}
(d H^2 + b H^3)^2\right]^{1/2} \nonumber \\ &&
- {1\over 4} \left\{ -{a\over H_0} + d{H_1\over H_0 H^3} 
+ b {H_1\over H_0 H^2} + {c\over H^1}\right\}\, .
\een
The equilibrium separation \refb{en2.14} between the 
two centers can be found by
extremizing \refb{en2.40} with respect to $r$.
Corresponding result for the second system is obtained by exchanging
$b$ and $d$ in \refb{en2.40}.

\sectiono{Replacing the enhancon by the smooth solution} \label{s2.2}

In this section we shall replace the solution in the S-T-U model described
in \S\ref{sdiag} by  a smooth dyon solution and compute the range
of values of $\zeta_1$ for which the solution exists. Since the analysis of this
section will be somewhat technical let us first summarize the main result. We shall
find that the net effect of smoothening the solution is to replace in the
expressions for $\Sigma$, $S_2$, $T$, $U$, $\AAA^{0}_0$, $\AAA^2_0$,
$\AAA^3_0$ and $\AAA_{10}$ given in \refb{en2.37},
the variable $r$ by $\hat r$ where
\be \label{ehatrdef}
{1\over \hat r} = {1\over r} - \kappa \coth (\kappa r) + \kappa, \qquad
\kappa \equiv \sqrt{\zeta_2\over 8\rho_2 \sigma_2} {|\rho_2 - \sigma_2|
\over |\zeta|} = {1\over r_e}\, .
\ee
This does not mean that the new solution is related to the old one by a
coordinate transformation since for example
the $dx^i dx^i$ term in the expression for
the metric is still given by $dr^2 + r^2 d\Omega_2^2$ with $d\Omega_2$
denoting the line element on a unit 2-sphere. Nevertheless it shows that the
potential for the test charge in this new background is given by \refb{en2.40}
with $r$ replaced by $\hat r$ everywhere. Thus for given values of the asymptotic
moduli the equilibrium position of the test charge $(Q,0)$ or $(Q+u P,0)$
is given by replacing $|\vec r_1-\vec r_2|=|\vec r_2|$ by $\hat r_2$
in the S-T-U model results \refb{en2.14} and \refb{en2.14aa} respectively, where
$\hat r_2$ is the value of $\hat r$ defined in \refb{ehatrdef} for
$r=|\vec r_2|$.\footnote{We are again setting $\vec r_1=0$ \i.e.\
taking the location of the first center as the origin of the coordinate system.} 
Now since according to \refb{ehatrdef} 
$r=0$ corresponds to $\hat r=r_e$ and $r=\infty$ corresponds to $\hat r=\infty$
we see that requiring $0 < |\vec r_2|<\infty$ corresponds to
$r_e < \hat r_2 < \infty$. 
This according to the analysis of \S\ref{s1} (with $r$ replaced by $\hat r$)
constraints $\tau$ to lie inside the region $R_1'$ of Fig.~\ref{f2} for the first
configuration and the region $R_2'$ of Fig.~\ref{f2} for the second
configuration.
Thus we
conclude that
the ranges of $\tau_1$ for which the solutions exists remain the same
as what we derived in \S\ref{s1}.
However the interpretation of what happens at the
left boundary of $R_1'$ and the right boundary of $R_2'$ is slightly different.
At the left boundary of $R_1'$,
when $\tau_1$ saturates the bound \refb{en2.25}, the second center of the first
configuration reaches the center of the smooth dyonic solution. 
On the other hand at the right boundary of $R_2'$, when
$\tau_1$ saturates the bound \refb{en2.30}, the second center of the second 
configuration reaches the center of the smooth dyonic solution. 

We shall now describe how these results arise.

\subsection{Harvey-Liu monopole and dyon solutions in the ten dimensional
description} \label{s2}

We shall consider a truncation of the effective
action of ten dimensional heterotic string theory where we keep only a single
$SU(2)$ gauge field $\VVV_\mu^{(a)}$ ($1\le a\le 3$) out of $SO(32)$ or
$E_8\times E_8$. This action
is given by
\ben \label{en0.1}
S &=& {2\pi \over (2\pi\sqrt{\alpha'})^8}\int d^{10}x \sqrt{-\det G} e^{-2\Phi}
\left[R + 4 G^{ M N}\p_ M\Phi\p_ N\Phi \right.\nonumber \\
&& \qquad \qquad \left. -{1\over 12}
G^{ M M'}G^{NN'}G^{ R R'} H_{ M N R}
H_{ M' N' R'} -{\alpha'\over 8} \WWW_{MN}^{(a)} \WWW^{(a)MN}\right]\, ,
\nonumber \\
\WWW^{(a)}_{MN} &\equiv& \p_M \VVV^{(a)}_N -   \p_N \VVV^{(a)}_M
+\epsilon^{abc} \VVV^{(b)}_M\VVV^{(c)}_N, 
\een
\be \label{en0.2}
dH = -{\alpha'\over 4} \WWW^{(a)} \wedge \WWW^{(a)},
\quad H \equiv {1\over 3!} H_{MNP} dx^M\wedge dx^N\wedge dx^P, \quad
\WWW^{(a)} \equiv {1\over 2!} \WWW^{(a)}_{MN} dx^M\wedge dx^N\,.
\ee
Here $x^M$ for $0\le M\le 9$ are the coordinates labelling the ten dimensional
space-time, $G_{ M N}$ is the string metric, $H$ is the 3-form field strength
and $\Phi$ is the
dilaton field. 
We now compactify the theory on $T^6$ labelled by
$x^4,\cdots x^9$ with period $2\pi\sqrt{\alpha'}$ and non-compact coordinates
labelled by $x^0,x^1,x^2,x^3$. In this theory we consider the Harvey-Liu 
monopole solution\cite{harliu,0110109}\footnote{Strictly speaking, if we
take the circles labelled by $x^6,\cdots x^9$ to have self-dual radius, as is the case
for the metric given in \refb{en0.3}, we shall get additional massless non-abelian
gauge fields. We can avoid this situation by taking the metric along the
$x^6,\cdots x^9$ direction to be $K_{mn} dx^m dx^n$ for some constant symmetric
matrix $K$ with $\det K=1$. This does not affect any of the subsequent analysis.
Similarly we could also break the rest of the ten dimensional gauge group 
(SO(32) or $E_8\times E_8$) by turning on Wilson lines for these
gauge fields along the 6-7-8-9 directions without changing any of the
subsequent analysis.}
\ben \label{en0.3}
\VVV^{(a)}_i &=& \eps_{iak}{x^k\over r^2} (K(C_1r)-1), \quad \VVV^{(a)}_4
= C_2 \, {x^a\over r^2} H(C_1r), \quad 1\le i,k,a\le 3, \quad r\equiv \sqrt{x^k x^k}\, ,
 \nonumber \\
&& H(x)\equiv x\coth x -1, \quad K(x) = x/\sinh x\, , \nonumber \\
e^{2\Phi} &=&  C_3^2 + {\alpha'\over 4} (C_1^2 - r^{-2} H(C_1 r)^2 ) \, ,
\nonumber \\
ds^2 &=& - (dx^0)^2 + e^{2\Phi} \left( (dx^1)^2 + (dx^2)^2 + (dx^3)^2 +
C_2^2 (dx^4)^2\right) + C_4^2 (dx^5)^2 + \sum_{m=6}^9 dx^m dx^m \, , \nonumber \\
H_{4ij} &=&  - 2 \,  C_2\, e^{2\Phi} \,  \epsilon_{ijk}\p_k \Phi \quad 1\le i,j,k\le 3\, ,
\een
where $C_1$, $C_2$, $C_3$ and $C_4$ are arbitrary constants, 
and $\eps_{ijk}$ is the totally anti-symmetric symbol with $\epsilon_{123}=1$.
Since all the fields in \refb{en0.3} are invariant under changes in signs
of $C_1$, $C_3$ and $C_4$, we can choose
\be\label{esignchoice}
C_1, C_4, C_2 C_3 > 0\, ,
\ee
without any loss of generality.
Note that the solution described in \refb{en0.3} lies outside the
truncated theory described in \S\ref{strunc} since we have non-trivial background values of the
ten dimensional gauge fields. However we shall see that (the
dyonic generalization of)
this solution can be mapped to a solution inside the truncated theory by a duality rotation.

Physically \refb{en0.3} represents a gravitationally dressed
BPS monopole solution of the $SU(2)$ gauge
theory.
We can construct
from this a dyon solution by making the replacement (see {\it e.g.} \cite{0609055})
\be \label{en0.4}
x^0 \to \cosh\gamma \, x^0 + C_2 C_3\sinh\gamma \, x^4, \quad
x^4 \to C_2^{-1} C_3^{-1}  \sinh\gamma \, x^0 + \cosh\gamma \, x^4\, ,
\ee
and taking the new $x^4$ coordinate defined this way as being periodically
identified with period $2\pi\sqrt{\alpha'}$. This gives a solution:
\ben \label{en0.5}
\VVV^{(a)}_i &=& \eps_{iak}{x^k\over r^2} (K(C_1r)-1), \quad \VVV^{(a)}_4
= C_2 \, \cosh\gamma \, 
{x^a\over r^2} H(C_1r) \, , \nonumber \\
\VVV^{(a)}_0
&=&  C_3^{-1}\, \sinh\gamma \, 
{x^a\over r^2} H(C_1r) \, , \nonumber \\
e^{2\Phi} &=&  C_3^2 + {\alpha'\over 4} (C_1^2 - r^{-2} H(C_1 r)^2 ) \, , \nonumber \\
ds^2 &=& - (dx^0)^2 + e^{2\Phi} \left( (dx^1)^2 + (dx^2)^2 + (dx^3)^2 
\right) + C_2^2 C_3^2 (dx^4)^2+ C_4^2 (dx^5)^2 + \sum_{m=6}^9 dx^m dx^m
\nonumber \\
&& + \left(e^{2\Phi}C_3^{-2}-1\right) (\sinh\gamma \, dx^0 + C_2 C_3\, \cosh\gamma \, dx^4)^2 \, ,
\nonumber \\
H_{4ij} &=& -2\, C_2\, \cosh\gamma \, e^{2\Phi} \, 
\epsilon_{ijk}\p_k \Phi\, ,
\nonumber \\
H_{0ij} &=& -2 \, C_3^{-1} \,
\sinh\gamma \, e^{2\Phi} \, \epsilon_{ijk}\p_k  \Phi, 
\quad 1\le i,j,k,a\le 3\, .
\een

The solutions given above are in the hedgehog  gauge. For comparison with the 
solution in the S-T-U model it will be more appropriate to express the solution
in the
string gauge (see {\it e.g.} \cite{0110109}). In this gauge the solution takes the form
\ben \label{en0.6}
\VVV^{(3)}_i dx^i  &\simeq & \cos\theta d\phi, \nonumber \\
\VVV^{(3)}_4
&=& C_2 \, \cosh\gamma \, 
{1\over r} H(C_1r)\, , \nonumber \\
\VVV^{(3)}_0
&=&  C_3^{-1}\, \sinh\gamma \, 
{1\over r} H(C_1r) \nonumber \\
e^{2\Phi} &=&  C_3^2 + {\alpha'\over 4} (C_1^2 - r^{-2} H(C_1 r)^2 ) \nonumber \\
ds^2 &=& - (dx^0)^2 + e^{2\Phi} \left( (dx^1)^2 + (dx^2)^2 + (dx^3)^2 
\right) + C_2^2 C_3^2 (dx^4)^2+ C_4^2 (dx^5)^2 + \sum_{m=6}^9 dx^i dx^i \nonumber \\
&& + \left(e^{2\Phi}C_3^{-2}-1\right) (\sinh\gamma \, dx^0 + C_2 C_3\, \cosh\gamma \, dx^4)^2 
\nonumber \\
H_{4ij} &=& -2\, C_2\, \cosh\gamma \, e^{2\Phi}\, 
\epsilon_{ijk} \p_k \Phi, 
\nonumber \\
H_{0ij} &=& -2 \, C_3^{-1}\, \sinh\gamma \, e^{2\Phi}\, 
\epsilon_{ijk}\p_k \Phi, \quad
1\le i,j,k\le 3\, .
\een
The $\simeq$ in the first equation describes equality up to terms of
order $e^{-C_1 r}$ and also additive constants. 

{}From now on we shall work in the $\alpha'=16$ unit.
For reason that will become clear later,
we shall choose the constants $C_i$'s and $\gamma$ such that 
\be \label{en0.7}
G_{44} + 4 (\VVV^{(3)}_4)^2 = C_2^2 C_3^2 + 4 C_1^2 C_2^2 \cosh^2\gamma=1
\, .
\ee

\subsection{Smooth dyon solution in the four dimensional description} \label{s0}

We now translate the above solution into a field configuration in an effective
four dimensional field theory.
For this we dimensionally reduce the theory to four dimensions, keeping 
a single $U(1)$ gauge field $\VVV^{(3)}_ M$
in ten dimensions, and setting the components
of various fields along $T^4$, labelled by the coordinates
$x^6,\cdots x^9$, to their background values given in \refb{en0.6}, 
and setting $\alpha'=16$. This
leads to an action
whose bosonic part is given by:
\ben \label{e1}
S&=&{1\over 32\pi} \int d^4 x\, \sqrt{-\det g}\,
\bigg[R - {1\over 2\, S_2^2} g^{\mu\nu}\p_\mu
S \p_\nu \bar S
- S_2 F^{(a)}_{\mu\nu}
(LML)_{ab} F^{(b)\mu\nu} + S_1 F^{(a)}_{\mu\nu}
L_{ab} \wt F^{(b)\mu\nu} \nonumber \\ &&
+ {1\over 8} g^{\mu\nu} \, Tr(\p_\mu M L \p_\nu M L)\bigg]
\, .
\een
Here $S=S_1+iS_2$ is a complex scalar field representing the heterotic
axion - dilaton
system, $F^{(a)}_{\mu\nu}
\equiv \p_\mu A^{(a)}_\nu - \p_\nu A^{(a)}_\mu$ for $1\le a\le 5$ are the
gauge field strengths associated with five $U(1)$ gauge fields $A_\mu^{(a)}$,
$\wt F_{\mu\nu}$ denotes the dual field strength of $F_{\mu\nu}$, 
$L$ is the $5\times 5$ matrix
\be \label{e2}
L =\pmatrix{0 & I_2 & \cr I_2 & 0 & \cr  &   & -1}\, ,
\ee
with $I_n$ denoting $n\times n$ identity matrix, and $M$ is a matrix valued scalar
field, satisfying
\be \label{e3}
MLM^T = L, \qquad M^T=M\, .
\ee
The precise relation between the fields appearing here and those in the
ten dimensional supergravity was given in \cite{9207016} and 
reviewed in \cite{9402002}.\footnote{In the convention of \cite{9402002} that
we shall use, $S$ corresponds to the field $\lambda$.}
We shall use the normalization convention of \cite{9402002}, keeping in
mind that $\VVV_\mu^{(3)}$ is related to the ten dimensional abelian gauge fields
$A_\mu^{(10)I}$ used in \cite{9402002} as $A_\mu^{(10)1} = 2\sqrt 2 \VVV_\mu^{(3)}$. 
In order to facilitate comparison with the fields of the S-T-U model as reviewed in
\S\ref{s1}, where the
normalization in front of the Einstein-Hilbert time is given by $1/16\pi$, we shall
make a $g_{\mu\nu}\to 2 g_{\mu\nu}$ field redefinition, so that the action takes
the form:
\ben \label{e1alt}
S&=&{1\over 16\pi} \int d^4 x\, \sqrt{-\det g}\,
\bigg[R - {1\over 2\, S_2^2} g^{\mu\nu}\p_\mu
S \p_\nu \bar S
- {1\over 2} S_2 F^{(a)}_{\mu\nu}
(LML)_{ab} F^{(b)\mu\nu} + {1\over 2} S_1 F^{(a)}_{\mu\nu}
L_{ab} \wt F^{(b)\mu\nu} \nonumber \\ &&
+ {1\over 8} g^{\mu\nu}\, Tr(\p_\mu M L \p_\nu M L)\bigg]
\, .
\een

If we denote the metric appearing in \refb{en0.6} by $G_{MN}$ and 
define 
\be \label{edefa4}
A_4 = 2\sqrt 2 \VVV^{(3)}_4,
\ee 
then 
using the results reviewed in \cite{9402002} we find that the four dimensional
field configuration corresponding to the background \refb{en0.6} is
given by\footnote{In order to get the expression for $S_1$ 
 given in \refb{es1result}, we need to correct the formula for the 4-dimensional
2-form field $B_{\mu\nu}$ given in eq.(3) of \cite{9402002}. The corrected expression
is given by
$$ B_{\mu\nu}=B^{(10)}_{\mu\nu} - 4 \wh B_{mn} A^{(m)}_\mu A^{(n)}_\nu  
- 2 \left( A^{(m)}_\mu A^{(m+6)}_\nu -A^{(m)}_\nu A^{(m+6)}_\mu\right) 
- 2 \wh A^I_m \left( A^{(I+12)}_\mu A^{(m)}_\nu - A^{(I+12)}_\nu A^{(m)}_\mu\right) $$
in the notation of \cite{9402002}. The last term
was missed in \cite{9402002} but is needed to ensure that $B_{\mu\nu}$ transforms
correctly under the gauge transformation of $A^{(I+12)}_\mu$.}
\ben \label{e30.1}
M &=& \pmatrix{G_{55}^{-1} & 0 & 0 & 0 & 0\cr
0 & G_{44}^{-1} & 0 & {1\over 2}G_{44}^{-1} A_4^2 & G_{44}^{-1} A_4 \cr 
0 & 0 & G_{55} & 0 & 0 \cr
0 & {1\over 2}G_{44}^{-1} A_4^2 & 0 & (G_{44}+{1\over 2}A_4^2)^2 G_{44}^{-1}
&  (G_{44}+{1\over 2}A_4^2) G_{44}^{-1} A_4 \cr
0 & G_{44}^{-1} A_4 & 0 &  (G_{44}+{1\over 2}A_4^2) G_{44}^{-1} A_4 &
1 +G_{44}^{-1} A_4^2}\nonumber \\
&=& \pmatrix{G_{55}^{-1} & 0 & 0 & 0 & 0\cr
0 & G_{44}^{-1} & 0 & {1\over 2}G_{44}^{-1} A_4^2 & G_{44}^{-1} A_4 \cr 
0 & 0 & G_{55} & 0 & 0 \cr
0 & {1\over 2}G_{44}^{-1} A_4^2 & 0 &  G_{44}^{-1}
&   G_{44}^{-1} A_4 \cr
0 & G_{44}^{-1} A_4 & 0 &  G_{44}^{-1} A_4 &
 1 +G_{44}^{-1} A_4^2} \, ,
\een
where in the last step we have used \refb{en0.7},
\be \label{e30.3}
S_2 = e^{-2\Phi} C_2 C_4\sqrt{e^{2\Phi} \cosh^2 \gamma - C_3^2 \sinh^2\gamma}\, ,
\ee
\be \label{es1result}
S_1 \simeq C_2 C_3 C_4 \, \sinh\gamma\, e^{-2\Phi} \, ,
\ee
\ben \label{e30.2}
\{A^{(a)}_0 \}
= - \sqrt 2 \, C_3\, \sinh\gamma \, \left(e^{2\Phi}\cosh^2\gamma 
-C_3^2 \sinh^2\gamma\right)^{-1}\, \,  \pmatrix{0\cr 
-{1\over 2\sqrt 2} C_2^{-1} \cosh\gamma (e^{2\Phi} C_3^{-2} -1)\cr 0\cr 
\sqrt{2} \, C_2 \, \cosh\gamma\, r^{-2} H(C_1 r)^2 \cr r^{-1} H(C_1 r)},
\nonumber \\ 
\quad \{ A^{(a)}_i dx^i\} = -  \sqrt 2\, \cos\theta \, d\phi \, 
\pmatrix{0\cr 0\cr 0\cr 0\cr 1}\, , \qquad \qquad \qquad \qquad 
\qquad \qquad \qquad \qquad \qquad \qquad \qquad \qquad 
\een
\be \label{e30.4}
g_{\mu\nu} dx^\mu dx^\nu = - {C_2C_4\over 2\sqrt{e^{2\Phi} \cosh^2 \gamma - C_3^2 \sinh^2\gamma}} (dx^0)^2 + {1\over 2} \, C_2 C_4 
\sqrt{e^{2\Phi} \cosh^2 \gamma - C_3^2 \sinh^2\gamma}
\, \, dx^i dx^i \, .
\ee

We now take the $5\times 5$ matrix
\ben \label{e2.0}
W &\equiv& \pmatrix{{I_2/\sqrt 2} & {I_2/ \sqrt 2} &
\cr {I_2/ \sqrt 2} & - {I_2/\sqrt 2} & \cr & & 1}
\pmatrix{I_3 & & \cr & 0 & 1\cr & 1 & 0} \pmatrix{{I_2/\sqrt 2} & {I_2/ \sqrt 2} &
\cr {I_2/ \sqrt 2} & - {I_2/\sqrt 2} & \cr & & 1}\nonumber \\
&=& \pmatrix{1 & 0 & 0 & 0 & 0\cr
0 & {1\over 2} & 0 & {1\over 2} & {1\over \sqrt 2}\cr
0 & 0 & 1 & 0 & 0\cr 0 & {1\over 2} & 0 & {1\over 2} & -{1\over \sqrt 2}\cr
0 & {1\over \sqrt 2} & 0 & - {1\over \sqrt 2} & 0
}
\, , 
\een
satisfying
\be \label{e2.01}
W^T W = I_5, \qquad W^T L W = L\, ,
\ee
and make the field redefinition:
\be \label{e2.2}
M\to W M W^T, \qquad F^{(a)}_{\mu \nu} \to W_{ab} F^{(b)}_{\mu \nu}\, .
\ee
The action in the new variables takes the same form as \refb{e1}.
After this transformation the solution \refb{e30.1} for $M$ becomes
\be \label{e2.02}
M = \pmatrix{\wt R^{-2} & 0 & 0 & 0 & 0\cr 0 & R^{-2} & 0 & 0 & 0\cr
0 & 0 & \wt R^2 & 0 & 0\cr 0 & 0 & 0 & R^2 & 0\cr 0 & 0 & 0 & 0 & 1}\, ,
\ee
where
\be \label{e2.03}
\wt R^2 = G_{55}=C_4^2, \quad 
R^2 = {1 - {1\over \sqrt 2} A_4 \over 1 + {1\over \sqrt 2} A_4}
= {1 - 2 \VVV^{(3)}_4\over 1 + 2 \VVV^{(3)}_4}
= {1 - 2 \, C_2\, \cosh\gamma  \, r^{-1} H(C_1 r) \over
1 + 2 \, C_2\, \cosh\gamma  \, r^{-1} H(C_1 r)}\, .
\ee
The gauge field background takes the form, up to constant shifts, 
\ben \label{e30.25}
\{A^{(a)}_0 \}
&=&  C_3 C_2^{2} \sinh\gamma \, {1\over r}\, H(C_1 r)\, \pmatrix{0\cr 
\left\{2C_2 \cosh\gamma\,
r^{-1}\, 
H(C_1 r) -1\right\}^{-1} \cr 0\cr \left\{2C_2 \cosh\gamma\,
r^{-1}\, 
H(C_1 r) +1\right\}^{-1} \cr 0}, \nonumber \\
\quad \{ A^{(a)}_i dx^i\} &\simeq& \cos\theta \, d\phi \, 
\pmatrix{0\cr -1\cr 0\cr 1\cr 0}\, .
\een
The metric and the axion-dilaton fields remain unchanged under this field redefinition.

We now note that for the solution described above the matrix $M$ and the 
gauge fields are non-trivial only along  
the first four rows and columns. This corresponds to
setting to zero all ten dimensional gauge fields and also setting all components
of the metric and 2-form fields with one or both legs along $T^4$ to trivial
values. This is precisely the condition under which the solution can be embedded
in the S-T-U model. 
Rescaling $x^i$ and $x^0$ as
\be \label{e40.1}
x^i \to \sqrt{2\over C_2 C_3 C_4} x^i, \quad x^0 \to x^0 \sqrt{2C_3\over  C_2 C_4}
\, ,
\ee
and
identifying $R\wt R$ with $T_2$ and $\wt R/ R$ with $U_2$ we see that
in the variables of the S-T-U model the scalar fields and the metric takes the form:
\ben \label{e40.2}
&& T_1 = 0, \quad U_1 = 0, \nonumber \\
&& T_2 U_2 = C_4^2, \quad {T_2\over U_2} = {1 -  C_2\, \cosh\gamma  \, 
\sqrt{2 \, C_2 C_3 C_4} \, r^{-1}  H(\sqrt 2\, C_1 r/\sqrt{C_2 C_3 C_4}) \over
1 + C_2\, \cosh\gamma  \, 
\sqrt{2 \, C_2 C_3 C_4} \, r^{-1}  H(\sqrt 2\, C_1 r/\sqrt{C_2 C_3 C_4})}\, ,
\nonumber \\
&& S_2 = e^{-2\Phi} C_2 C_4\sqrt{e^{2\Phi} \cosh^2 \gamma - C_3^2 \sinh^2\gamma}, 
\nonumber \\
&& S_1 \simeq C_2 C_3 C_4 \, \sinh\gamma\, e^{-2\Phi} \, ,
\nonumber \\
&&
g_{\mu\nu} dx^\mu dx^\nu = - e^{2V}\,  (dx^0)^2 + e^{-2V} \, \, dx^i dx^i \, , \nonumber \\
&& \quad 
e^{2\Phi} =  C_3^2 + 4 \left(C_1^2 - {C_2 C_3 C_4\over 2\, r^2}\,
H(\sqrt 2\, C_1 r/\sqrt{C_2 C_3 C_4})^2 \right),
\quad e^{2V} \equiv 
{C_3\over \sqrt{e^{2\Phi} \cosh^2 \gamma - C_3^2 \sinh^2\gamma}}  \, .
\nonumber \\
\een
To find the gauge fields in the S-T-U model notation we first note that after the
coordinate change  \refb{e40.1} the first four components of
gauge fields $A^{(a)}_\mu$ given in
\refb{e30.25} takes the form
\ben \label{egg1}
\{A^{(a)}_0 \}
&=&  C_3^2 C_2^{2} \sinh\gamma \, {1\over r}\, H(\sqrt 2\, C_1 r/\sqrt{C_2 C_3 C_4})
\nonumber \\
&& \pmatrix{0\cr 
\left\{2C_2 \cosh\gamma\, \sqrt{C_2 C_3 C_4\over 2}\, 
r^{-1}\, 
H(\sqrt 2\, C_1 r/\sqrt{C_2 C_3 C_4}) -1\right\}^{-1} \cr 0\cr \left\{2C_2 \cosh\gamma\,
\sqrt{C_2 C_3 C_4\over 2}\, r^{-1}\, 
H(\sqrt 2\, C_1 r/\sqrt{C_2 C_3 C_4}) +1\right\}^{-1}}, \nonumber \\
\quad \{ A^{(a)}_i dx^i\} &\simeq& \cos\theta \, d\phi \, 
\pmatrix{0\cr -1\cr 0\cr 1} \, .
\een
Now it was shown in \cite{9402002} that a test charge $(Q,0)$ couples to this
gauge field background through the action
\ben \label{egg2}
\pm {1\over 2} \int dx^\mu A^{(a)}_\mu  Q_a
&=&\pm {1\over 2} \int dx^\mu \left[ A^{(1)}_\mu \, Q_1 + A^{(2)}_\mu \, Q_2
+ A^{(3)}_\mu \, Q_3 + A^{(4)}_\mu \, Q_4 \right]\nonumber \\
&=& \pm {1\over 2} \int dx^\mu \left[ A^{(1)}_\mu \, q_0 +A^{(2)}_\mu \, q_3 
- A^{(3)}_\mu \, p^1 + A^{(4)}_\mu \, q_2
\right]\, .
\een
The $\pm$ sign reflects
the fact that the analysis of \cite{9402002} determines the normalization but
not the sign of the coupling of the gauge fields to the charges since the bosonic
action involving the $U(1)$ gauge fields 
has an $A_\mu\to - A_\mu$ symmetry. Comparing this with
\refb{ed1} we get
\be \label{ecomp}
\pmatrix{\AAA^0_\mu \cr \AAA^3_\mu \cr 
\AAA_{1\mu} \cr  \AAA^2_\mu }
= \pm \pmatrix{A^{(1)}_\mu\cr A^{(2)}_\mu\cr A^{(3)}_\mu\cr A^{(4)}_\mu}\, .
\ee
Eq.\refb{egg1} now shows that the magnetic part of the field is given by
\be \label{emagnetic}
\AAA^3_i \,  dx^i \simeq \mp \cos\theta d\phi, \quad \AAA^2_i \,  dx^i
=\pm \cos\theta dx^i\, .
\ee
On the other hand \refb{en2.37} shows that the expected magnetic field
in the S-T-U model, produced by the first center, is given by
\be \label{emagnetic2}
\AAA^3_i \,  dx^i = - \cos\theta d\phi, \quad \AAA^2_i \,  dx^i
=\cos\theta dx^i\, .
\ee
Comparing \refb{emagnetic} and \refb{emagnetic2} we see that we must use
the top sign in \refb{ecomp}. This can now be used to express the electric
potentials given in \refb{egg1} as
\ben \label{egg3}
\pmatrix{\AAA^0_0 \cr \AAA^3_0 \cr 
\AAA_{10} \cr  \AAA^2_0 }
= \pmatrix{A^{(1)}_0\cr A^{(2)}_0\cr A^{(3)}_0\cr A^{(4)}_0}
&=& C_3^2 C_2^{2} \sinh\gamma \, {1\over r}\, H\left(\sqrt 2\, C_1 r/\sqrt{C_2 C_3 C_4}
\right)
\nonumber \\
&& \pmatrix{0\cr 
\left\{2C_2 \cosh\gamma\, \sqrt{C_2 C_3 C_4\over 2}\, 
r^{-1}\, 
H(\sqrt 2\, C_1 r/\sqrt{C_2 C_3 C_4}) -1\right\}^{-1} \cr 0\cr \left\{2C_2 \cosh\gamma\,
\sqrt{C_2 C_3 C_4\over 2}\, r^{-1}\, 
H(\sqrt 2\, C_1 r/\sqrt{C_2 C_3 C_4}) +1\right\}^{-1} }\, . \nonumber \\
\een
Finally, adding constant terms to the
gauge potential, we can bring \refb{egg3} to the form:
\ben \label{egg3.55}
\pmatrix{\AAA^0_0 \cr \AAA^3_0 \cr 
\AAA_{10} \cr  \AAA^2_0 }
&=&  C_3^2 C_2^{2} \sinh\gamma \, {1\over 2 C_2 \cosh\gamma} 
\sqrt{2\over C_2 C_3 C_4} \nonumber \\
&& \pmatrix{0\cr 
\left\{2C_2 \cosh\gamma\, \sqrt{C_2 C_3 C_4\over 2}\, 
r^{-1}\, 
H(\sqrt 2\, C_1 r/\sqrt{C_2 C_3 C_4}) -1\right\}^{-1} \cr 0\cr -\left\{2C_2 \cosh\gamma\,
\sqrt{C_2 C_3 C_4\over 2}\, r^{-1}\, 
H(\sqrt 2\, C_1 r/\sqrt{C_2 C_3 C_4}) +1\right\}^{-1}} \, .
\een

Defining
\be \label{e40.7}
{1\over \hat r} = \kappa - {1\over r}\, H(\kappa r) =
 {1\over r} - \kappa \coth (\kappa r) + \kappa, 
\ee
where
\be \label{e40.8}
\kappa = {\sqrt 2\, C_1\over  \sqrt{C_2 C_3 C_4}}\, ,
\ee
we can express \refb{egg3.55}, \refb{e40.2} as
\ben \label{emagfin}
&& \pmatrix{\AAA^0_0 \cr \AAA^3_0 \cr 
\AAA_{10} \cr  \AAA^2_0 }
= {1\over 2} \, {C_3\over C_2 C_4} \, {\sinh\gamma\over
\cosh^2\gamma} \, \pmatrix{0 \cr \left\{ - {1-2 C_1 C_2 \cosh\gamma \over 2 C_1
C_2\cosh\gamma} \kappa - {1\over \hat r}\right\}^{-1} \cr 0\cr 
\left\{ - {1+2 C_1 C_2 \cosh\gamma \over 2 C_1
C_2\cosh\gamma} \kappa + {1\over \hat r}\right\}^{-1}} \, , \nonumber \\
&& T_1 = 0, \quad U_1 = 0, \nonumber \\
&& T_2 U_2 = C_4^2, \quad {T_2\over U_2} = {1 -  C_2\, \cosh\gamma  \, 
\sqrt{2 \, C_2 C_3 C_4} \, (\kappa - \hat r^{-1}) \over
1 + C_2\, \cosh\gamma  \, 
\sqrt{2 \, C_2 C_3 C_4} \, (\kappa - \hat r^{-1}) }
\nonumber \\
&& S_2 = e^{-2\Phi} C_2 C_4\sqrt{e^{2\Phi} \cosh^2 \gamma - C_3^2 \sinh^2\gamma}, 
\nonumber \\
&& S_1 \simeq C_2 C_3 C_4 \, \sinh\gamma\, e^{-2\Phi} \, ,
\nonumber \\
&&
g_{\mu\nu} dx^\mu dx^\nu = - e^{2V}\,  (dx^0)^2 + e^{-2V} \, \, dx^i dx^i \, , \nonumber \\
&&
e^{2\Phi} =  C_3^2 + 4 \left(C_1^2 - {C_2 C_3 C_4\over 2}\,
(\kappa - \hat r^{-1})^2 \right),
\quad e^{2V} \equiv 
{C_3\over \sqrt{e^{2\Phi} \cosh^2 \gamma - C_3^2 \sinh^2\gamma}}  \, .
\nonumber \\
\een

For large $r$ we have $H(r)\simeq r-1$ and hence $\hat r \simeq r$ up to
exponentially suppressed corrections. 
In that case the field configurations
given in \refb{emagfin} agree with those 
given in \refb{en2.37} (up to constant additive terms in the gauge potential)
with the choice
\ben \label{e40.3}
&& \rho_2 = C_4 \sqrt{1- 2 C_1 C_2 \cosh\gamma\over 1+ 2 C_1 C_2 \cosh\gamma}\, ,
\quad \sigma_2 = 
C_4 \sqrt{1+ 2 C_1 C_2 \cosh\gamma\over 1- 2 C_1 C_2 \cosh\gamma}\, ,
\nonumber \\
&& \zeta_2 = {C_2 C_4\over C_3}, \quad \zeta_1 =  {C_2 C_4\over C_3} \, 
\sinh\gamma\, .
\een
Under this identification, $\kappa$ given in \refb{e40.8} becomes
\be \label{ekappanew}
\kappa = \sqrt{\zeta_2\over 8 \rho_2\sigma_2} {|\rho_2-\sigma_2|\over |\zeta|} = {1\over r_e}\, .
\ee

Now note that for finite $r$ the solutions for $S_2$, $T$, $U$, $V$ and $\AAA^{I}_0$,
$\AAA_{I0}$ 
are given by the same expressions as in the case of S-T-U model described in
\S\ref{s1} with the replacement of $r$ by $\hat r$.
Since these are the fields which determine  the location of the  
test particle charge (by the extrema of
\refb{en2.40}), we can directly take the results of section \ref{s1} with $r$ replaced by
$\hat r$ for determining the location of the test charge. 
Now from \refb{e40.7} we see that the condition $r>0$ corresponds to
$\hat r > 1/\kappa = r_e$.  Thus requiring $|\vec r_2|$ to be
positive corresponds to requiring $\hat r_2$, -- the value of $\hat r$ corresponding
to the vector $\vec r_2$ --  be larger that $r_e$. On the other hand for large $r$
we have $r\simeq \hat r$. Thus the condition $0<|\vec r_2|<\infty$ translates
to $r_e\le \hat r_2 < \infty$. Since we can use the results of \S\ref{s1} for determining
the location of the test charge with $r$ replaced by $\hat r$,  we see that
the condition $r_e\le \hat r_2 < \infty$ translates to requiring
$\tau_1$ to lie inside the range given in 
 \refb{en2.23a}, \refb{en2.25} for the first configuration and inside the
range given in \refb{en2.29}, \refb{en2.30} for the second configuration. 
These two ranges do not overlap, and 
together they make up the region $R_1'\cup R_2'$ 
of the moduli space shown in Fig.~\ref{f2}
-- precisely in agreement with the microscopic result for the index.

This still leaves open the question as to how the two configurations metamorphose
into each other at the boundary $L$ of $R_1'$ and $R_2'$. To examine this we apply
the inverse of the duality transformation \refb{e2.0} to map the test electric charges
$Q=\pmatrix{a\cr b\cr c\cr d\cr 0}$ and $Q+uP=\pmatrix{a\cr d\cr c\cr b\cr 0}$ to
\be \label{enewcharge}
\pmatrix{a\cr (b+d)/2 \cr c\cr (b+d)/2 \cr (b-d)/\sqrt{2}} \quad \hbox{and}
\quad \pmatrix{a\cr (b+d)/2 \cr c\cr (b+d)/2 \cr (d-b)/\sqrt{2}}\, .
\ee
The last entry represents electric charge under the $T^3$ generator of the SU(2)
group.
Now at the center of the dyon solution
the SU(2) gauge symmetry is restored. Thus at no cost in energy,
the test electric charge
can undergo an SU(2) rotation of $\pi$ about the 1-axis flipping the sign of the $T^3$
charge. This exchanges the quantum number $b$ and $d$, precisely
transforming the test electric charges of the two configurations to each other. 
Thus we see that the
two configurations can transform into each other at the boundary $L$ between
$R_1'$ and $R_2'$. The excess charge $-uP$ is dumped into the background, but
we do not detect it in the probe approximation that we are using.

\bigskip

{\bf Acknowledgement:}
We would like to thank Rajesh Gopakumar and Dileep Jatkar for useful
discussions.
This work was
supported in part by  the 
project 11-R\&D-HRI-5.02-0304. 
SL would like to thank the Harish-Chandra Research Institute for support 
in the form of a Senior Research Fellowship while part of the work was 
carried out.
The work of A.~Sen was also supported
in part by the J. C. Bose fellowship of 
the Department of Science and Technology, India.
Finally we would like to thank the people of India for their 
generous support to research in theoretical science.


\small

\baselineskip 12pt

\begin{thebibliography}{99}

\bibitem{9903163}
J.~Maldacena,  G.~Moore and A.~Strominger,
``Counting BPS blackholes in toroidal type II string theory,''
arXiv:hep-th/9903163.

\bibitem{0506151}
  D.~Shih, A.~Strominger and X.~Yin,
  ``Counting dyons in N = 8 string theory,''
  JHEP {\bf 0606}, 037 (2006)
  [arXiv:hep-th/0506151].


\bibitem{0803.1014}
  A.~Sen,
  ``N=8 Dyon Partition Function and Walls of Marginal Stability,''
  JHEP {\bf 0807}, 118 (2008)
  [arXiv:0803.1014 [hep-th]].


\bibitem{9607026}
R.~Dijkgraaf, E.~P.~Verlinde and H.~L.~Verlinde,
``Counting dyons in N = 4 string theory,''
Nucl.\ Phys.\ B {\bf 484}, 543 (1997)
[arXiv:hep-th/9607026].

\bibitem{0412287}
  G.~Lopes Cardoso, B.~de Wit, J.~Kappeli and T.~Mohaupt,
  ``Asymptotic degeneracy of dyonic 
  N = 4 string states and black hole
  entropy,''
  JHEP {\bf 0412}, 075 (2004)
  [arXiv:hep-th/0412287].

\bibitem{0505094}
D.~Shih, A.~Strominger and X.~Yin,
``Recounting dyons in N = 4 string theory,''
arXiv:hep-th/0505094.


\bibitem{0506249}
D.~Gaiotto,
``Re-recounting dyons in N = 4 string theory,''
arXiv:hep-th/0506249.

\bibitem{0508174}
  D.~Shih and X.~Yin,
  ``Exact black hole degeneracies and the topological string,''
  JHEP {\bf 0604}, 034 (2006)
  [arXiv:hep-th/0508174].

\bibitem{0510147}
  D.~P.~Jatkar and A.~Sen,
  ``Dyon spectrum in CHL models,''
  JHEP {\bf 0604}, 018 (2006)
  [arXiv:hep-th/0510147].

\bibitem{0605210}
  J.~R.~David and A.~Sen,
  ``CHL dyons and statistical entropy function from D1-D5 system,''
  JHEP {\bf 0611}, 072 (2006)
  [arXiv:hep-th/0605210].

\bibitem{0607155}
  J.~R.~David, D.~P.~Jatkar and A.~Sen,
  ``Dyon spectrum in N = 4 supersymmetric type II string theories,''
  arXiv:hep-th/0607155.


\bibitem{0609109}
  J.~R.~David, D.~P.~Jatkar and A.~Sen,
  ``Dyon spectrum in generic N = 4 supersymmetric Z(N) orbifolds,''
  arXiv:hep-th/0609109.

\bibitem{0802.0544}
  S.~Banerjee, A.~Sen and Y.~K.~Srivastava,
  ``Generalities of Quarter BPS Dyon 
Partition Function and Dyons of Torsion
  Two,''
  arXiv:0802.0544 [hep-th].

\bibitem{0802.1556}
  S.~Banerjee, A.~Sen and Y.~K.~Srivastava,
  ``Partition Functions of Torsion $>1$ Dyons in Heterotic
String Theory on $T^6$,''
  arXiv:0802.1556 [hep-th].

\bibitem{0803.2692}
  A.~Dabholkar, J.~Gomes and S.~Murthy,
  ``Counting all dyons in N =4 string theory,''
  arXiv:0803.2692 [hep-th].


\bibitem{0702141}
  A.~Sen,
  ``Walls of marginal stability and dyon spectrum in N = 4 supersymmetric
  string theories,''
  arXiv:hep-th/0702141.

\bibitem{0702150}
  A.~Dabholkar, D.~Gaiotto and S.~Nampuri,
  ``Comments on the spectrum of CHL dyons,''
  arXiv:hep-th/0702150.
  
\bibitem{0706.2363}
  M.~C.~N.~Cheng and E.~Verlinde,
  ``Dying Dyons Don't Count,''
  arXiv:0706.2363 [hep-th].


\bibitem{1104.1498} 
  A.~Sen,
  ``Negative discriminant states in N=4 supersymmetric string theories,''
  JHEP {\bf 1110}, 073 (2011)
  [arXiv:1104.1498 [hep-th]].


\bibitem{0712.3625}
  K.~Narayan,
  ``On the internal structure of dyons 
  in N = 4 super Yang-Mills theories,''
  Phys.\ Rev.\  {\bf D77}, 046004 (2008).
  [arXiv:0712.3625 [hep-th]].

\bibitem{0005049}
  F.~Denef,
  ``Supergravity flows and D-brane stability,''
  JHEP {\bf 0008}, 050 (2000)
  [arXiv:hep-th/0005049].

\bibitem{0304094} 
  B.~Bates and F.~Denef,
  ``Exact solutions for supersymmetric stationary black hole composites,''
  JHEP {\bf 1111}, 127 (2011)
  [hep-th/0304094].

\bibitem{0708.1270} 
  A.~Sen,
  ``Black Hole Entropy Function, Attractors and Precision Counting of Microstates,''
  Gen.\ Rel.\ Grav.\  {\bf 40}, 2249 (2008)
  [arXiv:0708.1270 [hep-th]].

\bibitem{9507090}
  M.~Cvetic and D.~Youm,
  ``Dyonic BPS saturated black holes of heterotic string on a six torus,''
  Phys.\ Rev.\ D {\bf 53}, 584 (1996)
  [arXiv:hep-th/9507090].
 
 \bibitem{9508094}
  M.~J.~Duff, J.~T.~Liu and J.~Rahmfeld,
  ``Four-Dimensional String-String-String Triality,''
  Nucl.\ Phys.\ B {\bf 459}, 125 (1996)
  [arXiv:hep-th/9508094].


\bibitem{9911161} 
  C.~V.~Johnson, A.~W.~Peet and J.~Polchinski,
  ``Gauge theory and the excision of repulson singularities,''
  Phys.\ Rev.\ D {\bf 61}, 086001 (2000)
  [hep-th/9911161].

\bibitem{harliu} 
  J.~A.~Harvey and J.~Liu,
  ``Magnetic monopoles in N=4 supersymmetric low-energy superstring theory,''
  Phys.\ Lett.\ B {\bf 268}, 40 (1991).

\bibitem{0110109} 
  M.~Wijnholt and S.~Zhukov,
  ``Inside an enhancon: Monopoles and dual Yang-Mills theory,''
  Nucl.\ Phys.\ B {\bf 639}, 343 (2002)
  [hep-th/0110109].

\bibitem{0609055} 
  E.~J.~Weinberg and P.~Yi,
  ``Magnetic Monopole Dynamics, Supersymmetry, and Duality,''
  Phys.\ Rept.\  {\bf 438}, 65 (2007)
  [hep-th/0609055].


\bibitem{9207016} 
  J.~Maharana and J.~H.~Schwarz,
  ``Noncompact symmetries in string theory,''
  Nucl.\ Phys.\ B {\bf 390}, 3 (1993)
  [hep-th/9207016].

\bibitem{9402002} 
  A.~Sen,
  ``Strong - weak coupling duality in four-dimensional string theory,''
  Int.\ J.\ Mod.\ Phys.\ A {\bf 9}, 3707 (1994)
  [hep-th/9402002].


\end{thebibliography}
\end{document}